\documentclass[reprint, superscriptaddress, amsmath, amssymb, aps, pra, a4paper]{revtex4-2}
\usepackage{graphicx}
\usepackage{dcolumn}
\usepackage{hyperref}
\usepackage{multirow}
\usepackage{comment}
\usepackage{xcolor}
\usepackage{amsmath}
\usepackage{mathrsfs}
\usepackage{mathtools}

\usepackage{caption}
\usepackage{subcaption}
\usepackage{natbib}
\usepackage{physics}
\usepackage{tabularx}
\usepackage{soul}

\begin{document}

\title{Mitigating the source-side channel vulnerability by characterization of photon statistics}

\author{Tanya Sharma}
\email{tanya@prl.res.in}
\affiliation{Quantum Science and Technology Laboratory, Physical Research Laboratory, Ahmedabad, India 380009.}
\affiliation{Indian Institute of Technology, Gandhinagar, India 382355.}

\author{Ayan Biswas}
\affiliation{School of Physics, Engineering $\&$ Technology and York Centre for Quantum Technologies, Institute for Safe Autonomy, University of York, YO10 5FT York, United Kingdom}

\author{Jayanth Ramakrishnan}
\affiliation{Quantum Science and Technology Laboratory, Physical Research Laboratory, Ahmedabad, India 380009.}

\author{Pooja Chandravanshi}
\affiliation{Quantum Science and Technology Laboratory, Physical Research Laboratory, Ahmedabad, India 380009.}

\author{Ravindra P. Singh}
\email{rpsingh@prl.res.in}
\affiliation{Quantum Science and Technology Laboratory, Physical Research Laboratory, Ahmedabad, India 380009.}

\date{\today}

\begin{abstract}

Quantum key distribution (QKD) theoretically offers unconditional security. Unfortunately, the gap between theory and practice threatens side-channel attacks on practical QKD systems. Many well-known QKD protocols use weak coherent laser pulses to encode the quantum information. These sources differ from ideal single photon sources and follow Poisson statistics. Many protocols, such as decoy state and coincidence detection protocols, rely on monitoring the photon statistics to detect any information leakage. The accurate measurement and characterization of photon statistics enable the detection of adversarial attacks and the estimation of secure key rates, strengthening the overall security of the QKD system. We have rigorously characterized our source to estimate the mean photon number employing multiple detectors for comparison against measurements made with a single detector. Furthermore, we have also studied intensity fluctuations to help identify and mitigate any potential information leakage due to state preparation flaws. We aim to bridge the gap between theory and practice to achieve information-theoretic security.
\end{abstract}


\maketitle

\section{\label{sec:Introduction}Introduction}

The advent of quantum computers significantly threatens existing classical cryptographic algorithms, and we can no longer guarantee the security of sensitive information with classical communication. QKD protocols \cite{BB84, E91, BBM92, DPSK, COW} have emerged as a promising approach to circumvent potential security concerns. 

However, practical implementations involve using imperfect devices that may assist an adversary in gaining partial information. The device imperfections may lead to further security threats by an adversary who may exploit such loopholes \cite{Huttner1994, Brassard2000, Vakhitov2001, Makarov2005, Gisin2006, Lydersen2010, Jain2011}. Characterising these devices is a crucial step towards ensuring the effectiveness of QKD in real-world scenarios.

One crucial aspect of optical quantum cryptography is using single-photon Fock states. While the experimental realisation of true single-photon sources remains challenging, many practical options, such as weak coherent pulses (WCPs), heralded single-photon sources, and the entangled single-photon source, are employed \cite{Xu2019}. 

Many implementations involve the use of WCPs as an approximation of single-photon Fock states. These WCPs are realised by highly attenuating a pulsed laser source using calibrated attenuators. It is a principle belief that laser sources operating well above the threshold emit a coherent state. Consequently, the faint laser pulses derived from such sources exhibit Poisson statistics regarding the number of photons per pulse. The mean photon number is the exclusive defining feature of Poissonian statistics. Therefore, ensuring the accurate measurement of the mean photon number is crucial for achieving secure QKD using a weak coherent source.

The Poisson statistics of the WCP source render a non-zero probability, however low, to attain more than a single photon per pulse. Such a loophole exposes our QKD system to adversarial attacks like the photon number splitting attack. The decoy state protocol \cite{Hwang2003, Lo2005, Wang2005, Ma2005}, involving the use of decoy pulses with slightly different mean photon numbers, has been proposed to mitigate this vulnerability. These decoy pulses are also characterised by Poisson statistics but with a slightly different mean photon number. Accurate estimation and characterisation of photon statistics, including the mean photon number, are crucial for implementing secure QKD protocols and detecting potential infiltration.

The practical implementation of QKD protocols heavily depends on the extensive utilisation of single-photon avalanche photodiodes (SPADs)\,\cite{Cova1981}. When evaluating the capabilities of a single-photon detector, it is essential to consider its spectral range, dead time, dark count rate, detection efficiency, timing jitters, and capacity to discern photon numbers \cite{Cheng2023}. The SPADs are threshold detectors which cannot provide information about the exact number of photons per pulse. They are also known as on-off detectors since they can only detect the presence or absence of a pulse containing photons. Hence using a single SPAD for characterising the source will give inaccurate estimates. Many studies \cite{Sperling2012, Sperling2013, Dynes2018, Kumazawa:19} discuss various approaches to photon characterisation using multiple on-off detectors. 

In this study, we focus on characterising the photon statistics of WCPs for QKD applications. Initially, we estimate the average photon number using a single detector. Subsequently, we employ a setup with four detectors to achieve a more accurate estimation of the average photon number. In a previous study \cite{Kumazawa:19}, the authors have introduced the utilisation of four detectors to accurately characterise the photon statistics of pulses by providing accurate probabilities for lower photon numbers $(n \leq 3)$. We adopt this characterisation approach to determine the Poisson statistics of the WCPs and obtain highly accurate estimations of the mean photon numbers. By comparing these results with previous estimates, we investigate the deviation of the mean photon number $(\mu)$ and analyse the resulting information leakage caused by this miscalculation.

Several investigations have delved into diverse state preparation deficiencies, offering security proofs that establish safeguarding measures even in the presence of these flaws \cite{Tamaki2014, Mizutani2015, Tamaki_2016, Nagamatsu2016, Wang2016, Wang2018, Pereira2019}. Lasers inevitably exhibit statistical fluctuations, resulting in inherent variability in the mean photon number they emit. However, it's crucial to quantify the magnitude of these fluctuations and comprehend their potential to facilitate information leakage. It is essential to recognise how the distinguishability of these fluctuations among different sources impacts the security of our Quantum Key Distribution (QKD) process. Furthermore, we have examined the fluctuations in intensity as a function of the source's intensity.

The organisation of this paper is as follows: We discuss the theoretical background to estimate the mean photon number per pulse and the fluctuations of a WCP source in Sec.\,\ref{sec:Theory}. We then discuss the experimental setup and procedures in Sec.\,\ref{sec:Experiment}. In Sec.\,\ref{sec:RnD}, we present the results for our source characterisation of intensity and fluctuations. Finally, we conclude and summarise our paper in Sec.\,\ref{sec:conc}.

\section{\label{sec:Theory}Theoretical Background}
\subsection{Coherent State}
The coherent state is specified in the Dirac notation as $\ket{\alpha}$, where $\alpha$ is the complex field amplitude. The coherent state is represented in terms of photon number states:

\begin{equation}\label{eq1}
    \ket{\alpha} = \mathrm{e}^{-|\alpha|^2 / 2} \sum^\infty_{n=0} \frac{\alpha^n}{\sqrt{n!}} \ket{n},
\end{equation}`

To find the probability $\mathcal{P}(n)$ that there are $n$ photons in the coherent state we evaluate $\braket{n}{\alpha}$ :
\begin{equation}\label{eq2}
    \begin{aligned}
\braket{n}{\alpha} & =\mathrm{e}^{-|\alpha|^2 / 2} \sum_{m=0}^{\infty} \frac{\alpha^m}{\sqrt{m!}}\braket{n}{m} \\
& =\mathrm{e}^{-|\alpha|^2 / 2} \sum_{m=0}^{\infty} \frac{\alpha^m}{\sqrt{m!}} \delta_{n m} \\
& =\mathrm{e}^{-|\alpha|^2 / 2} \frac{\alpha^n}{\sqrt{n!}},
\end{aligned}
\end{equation}

where, $\braket{n}{m} = \delta_{n,m}$, since number states are orthonormal. On equating $\mathcal{P}(n)$ with $|\braket{n}{\alpha}|^2$, we find:
\begin{equation}\label{eq3}
    \mathcal{P}(n) \equiv |\braket{n}{\alpha}|^2=\mathrm{e}^{-|\alpha|^2} \frac{\left|\alpha^2\right|^n}{n !} .
\end{equation}
\\
Taking, $|\alpha|^2=\mu$, we obtain a Poisson distribution $\mathcal{P}(n,\mu)$:
\begin{equation}\label{eq4}
    \mathcal{P}(n,\mu)=\frac{\mu^n}{n !} \mathrm{e}^{-\mu}.
\end{equation}

This establishes that coherent states have Poissonian photon statistics with mean photon number $\mu$.

\subsection{Weak Coherent Pulses (WCPs)}
A pulsed laser of repetition rate $\nu_{rep}$ and wavelength $\lambda$ is attenuated to generate WCPs. Since a laser is a coherent source, a faint laser source also emits coherent states. The number of photons per pulse is not deterministic but follows the Poisson distribution. Here, we discuss how we generate the WCPs of a desired distribution.

It is necessary for QKD sources to have a mean photon number of less than one to meet the security requirements of QKD protocols. We start by measuring the average power $P_{avg}$ of the source, from which we evaluate the energy per pulse $E_{pulse}$ as, 
\begin{equation}\label{eq5}
    E_{pulse} = \frac{P_{avg}}{\nu_{rep}}.
\end{equation}

The average number of photons per pulse $n_{avg}$ is given by, 

\begin{equation}\label{eq6}
    n_{avg} =\frac{E_{pulse} \lambda}{h c}.
\end{equation}

We utilize a neutral density filter with a defined optical density (OD) to attain the target mean photon number. The OD governs the degree of attenuation applied to the laser beam. Through the choice of an appropriate filter and varying attenuation, we can create WCPs exhibiting the desired Poisson statistics, characterized by a mean photon number denoted as $\mu_{desired}$, suitable for Quantum Key Distribution (QKD) applications.

\begin{equation}\label{eq7}
     \mu_{desired} = \frac{E_{pulse} \lambda}{hc}*10^{-OD}.
\end{equation}

\subsection{Estimations}
We discuss methods to estimate the mean photon number ($\mu$) of the source used in QKD implementation.
\subsubsection{Method-I : Using single detection}\label{M1}
Single photon detectors based on avalanche photodiodes are used to detect the number of detection per second. 
\begin{equation}\label{eq8}
    N = \mu* \nu_{rep} *\eta.
\end{equation}
Where $N$ is the number of detections per second, $\nu_{rep}$ is the repetition rate of the laser, and $\eta$ is detection efficiency. We can estimate the value of $\mu$ as, 

\begin{equation}\label{eq9}
    \mu = \frac{N}{\nu_{rep} *\eta}.
\end{equation}

Avalanche photodiodes are the prevailing choice for quantum signal detection, capable of indicating the presence or absence of photons within a pulse. However, relying on a single on-off detector could lead to underestimating photon statistics. Therefore, using photon-resolving detectors or a more comprehensive methodology becomes imperative in characterizing the QKD source. Accurate estimation of $\mu$ assumes paramount significance for accurate key rate calculations.

\subsubsection{Method-II Rigorous Characterisation}\label{M2}
\begin{figure}[htp]
    \centering
    \includegraphics[scale=0.45]{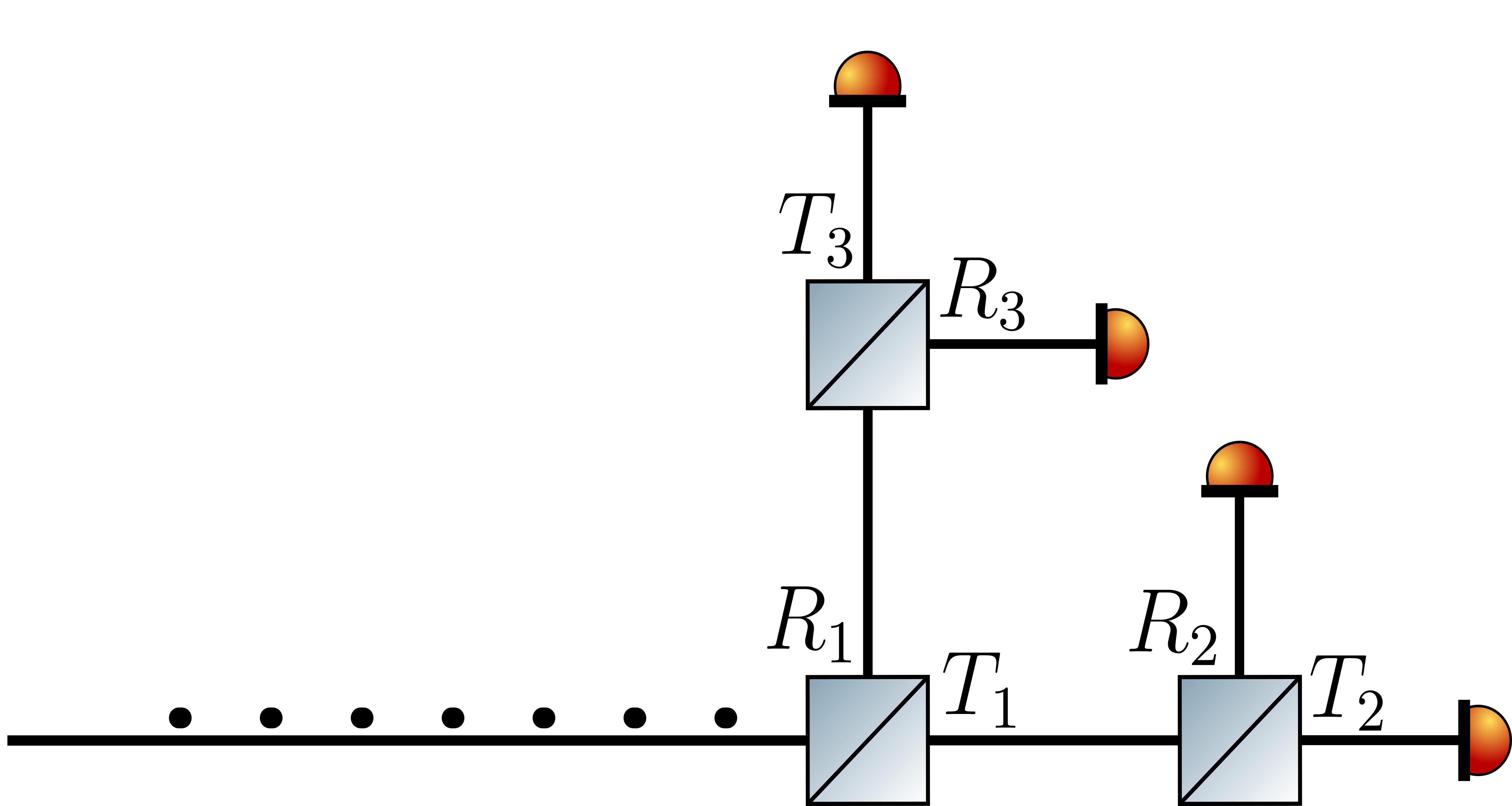}
    \caption{The branching efficiencies are defined as the probability for a photon to reach a particular detector. In the setup shown above the four branching efficiencies can be given as $\{T_1 T_2, T_1 R_2, R_1 T_3, R_1 R_3\}$.}
    \label{fig:branching}
\end{figure}
For source characterisation, splitting the pulse using an infinite number of beam splitters and threshold detectors is essential. The mean photon number in QKD is quite low; hence, we assume that multi-photon probability is minimal. This assumption enables us to rigorously characterise the source using four detectors ($D=4$). Figure \ref{fig:branching} illustrates an example of the setup necessary for conducting such comprehensive characterisation. The branching efficiencies define the probability of a photon reaching a specific detector. Let the branching efficiencies be $\eta_{b,i}$, the coupling efficiency be $\eta_c$, and the quantum efficiency of the detectors be $\eta_d$.

Overall efficiency is given as,
\begin{equation}\label{eq10}
\eta_i = \eta_{b,i}*\eta_c*\eta_d,   
\end{equation}
where $i=1,2,3,4$ represents all four arms and respective detectors. The average efficiency is given as,
\begin{equation}\label{eq11}
    \eta  := \frac{1}{4}\sum_{i=1}^D \eta_i.
\end{equation}

We define $Z_D:= {1,2,3,4}$ as a set of all the detectors. We want to record the r fold coincidences $(r=2,3,4)$, where $r=1$ refers to the counts in a single detector. Let us denote the observed r fold coincidence probability as $c_{obs,r}$ \cite{Kumazawa:19}.
\begin{equation}\label{eq12}
    c_{obs,r} = {\binom{D}{r}}^{-1} \sum_{W \subset I_r} c_{obs,W},
\end{equation}
where, $I_r:= \{W \subset Z_D\  \big| \ |W|=r\}$, $W$ is a subset of $Z_D$ with cardinality $|W|$, $c_{obs, W}$ denote the total coincidence probability of set $W$, where all $|W|$ detectors detect irrespective of the detection events in remaining detectors. 
The averaged r-fold coincidences, given the pulse, has $n$ photons, is,
\begin{equation}\label{eq13}
    c_{obs,r} = c_{n,r} \coloneqq \sum^{r}_{j=0} (-1)^j \omega_{r,j} \sum_{W \in I_j} (1-\sum_{i\in W}\eta_i)^n,
\end{equation}
where, 
\begin{equation}\label{eq14}
    \omega_{r,j} := \frac {\binom{D-j}{r-j}}{\binom{D}{r}}.
\end{equation}
For a Poissonian distribution with mean photon number $\mu$, the averaged $r(1,2,3,4)$ fold coincidences should satisfy,
\begin{equation}\label{eq15}
    c_{obs,r} = \sum_{n=0}^{\infty} p_n c_{n,r}, 
\end{equation}
where,
\begin{equation}\label{eq16}
    p_n = \frac{e^{-\mu} \mu^n}{n!}.
\end{equation}

To evaluate the coincidences from experimental data, we will employ equation \eqref{eq12} as referenced in our work. Additionally, we will utilize the bounds specified in the article \cite{Kumazawa:19} to experimentally validate the Poissonian statistics of the WCPs employed in QKD implementations and accurately estimate the mean photon number for the distribution.

\subsection{Information Leakage}
In principle, an eavesdropper may gain complete information on the bit value encoded on a multi-photon pulse. Privacy amplification enables us to reduce the information an eavesdropper might earn on the shared key between the authenticated parties. It is essential to determine the optimum subtraction bits for privacy amplification and avoid underestimating them. Since we cannot know Eve's attack strategy in advance, we consider an attack where Eve can access all the information in the multi-photon pulses. We could ensure absolute security by applying sufficient privacy amplification discarding all the bit values associated with the multi-photon pulses. If a WCP source has Poisson distribution with $\mu$ as the mean photon number per pulse, then the probability of multi-photons per pulse is given as,

\begin{equation}\label{eq17}
    p_{multi} = \sum_{p_n\geq 2} p_n
\end{equation}

While estimating the secure key rate, we consider this worst-case scenario, discard the coincidences, and consider only the single detection with a compatible basis. We estimate the subtraction terms as the multi-photon pulses that contribute to the single detections to consider the information leakage. Hence, we consider,

\begin{equation}\label{eq18}
    I(A:E) = \sum_{p_n\geq 2} \Bigl(p_n *\frac{1}{2^n}\Bigr)
\end{equation}

Where $\frac{1}{2^n}$ is the probability that a pulse containing n photons gives a single detection in the correct basis.

We encode using one degree of freedom, specifically polarization in our case. Any other degree of freedom associated with our states that could be used to discriminate the states could lead to a potential side-channel attack. Our objective is to investigate the differences in the fluctuations exhibited by the sources. When $N_s$ signal states with an intensity of $\mu$ are emitted, we posit that precisely $N_s \mu e^{-\mu}$ out of the $N_s$ signal states correspond to single photons. In practical situations, the quantity $\mu e^{-\mu}$ represents the probability of a single photon in the pulse. The observed count of single-photon signals will exhibit statistical fluctuations. We employ the methodology proposed in \cite{Biswas2021} to quantitatively assess the extent of information leakage to Eve. The correlation function $R$ characterizes the similarity between two sources. We compare $R$ with Eve's guessing probability $p(e|b)$, allowing the calculation of mutual information in terms of cross-correlation:
\begin{equation}\label{eq:8}
    I'(A:E) = 1 + \sum_{\substack{i,j \\ i\neq j}} \frac{R_{ij}}{4} \log_2 \left(\frac{R_{ij}}{4}\right).
\end{equation}

Here, $R_{ij}$ signifies the cross-correlation between sources $i$ and $j$. For perfectly identical functions, $I'(A:E)$ becomes zero, as indicated by Equation \ref{eq:8}. The measurement of cross-correlation offers an estimation of the extent of information leakage. Please note that to differentiate the information leakages due to multi-photon pulses and the information leakage due to side channels, we represent them by $I(A:E)$ and $I'(A:E)$, respectively.

\section{\label{sec:Experiment}Experimental Method }

\begin{figure*}[htp]
    \centering
    \includegraphics[width= 15cm]{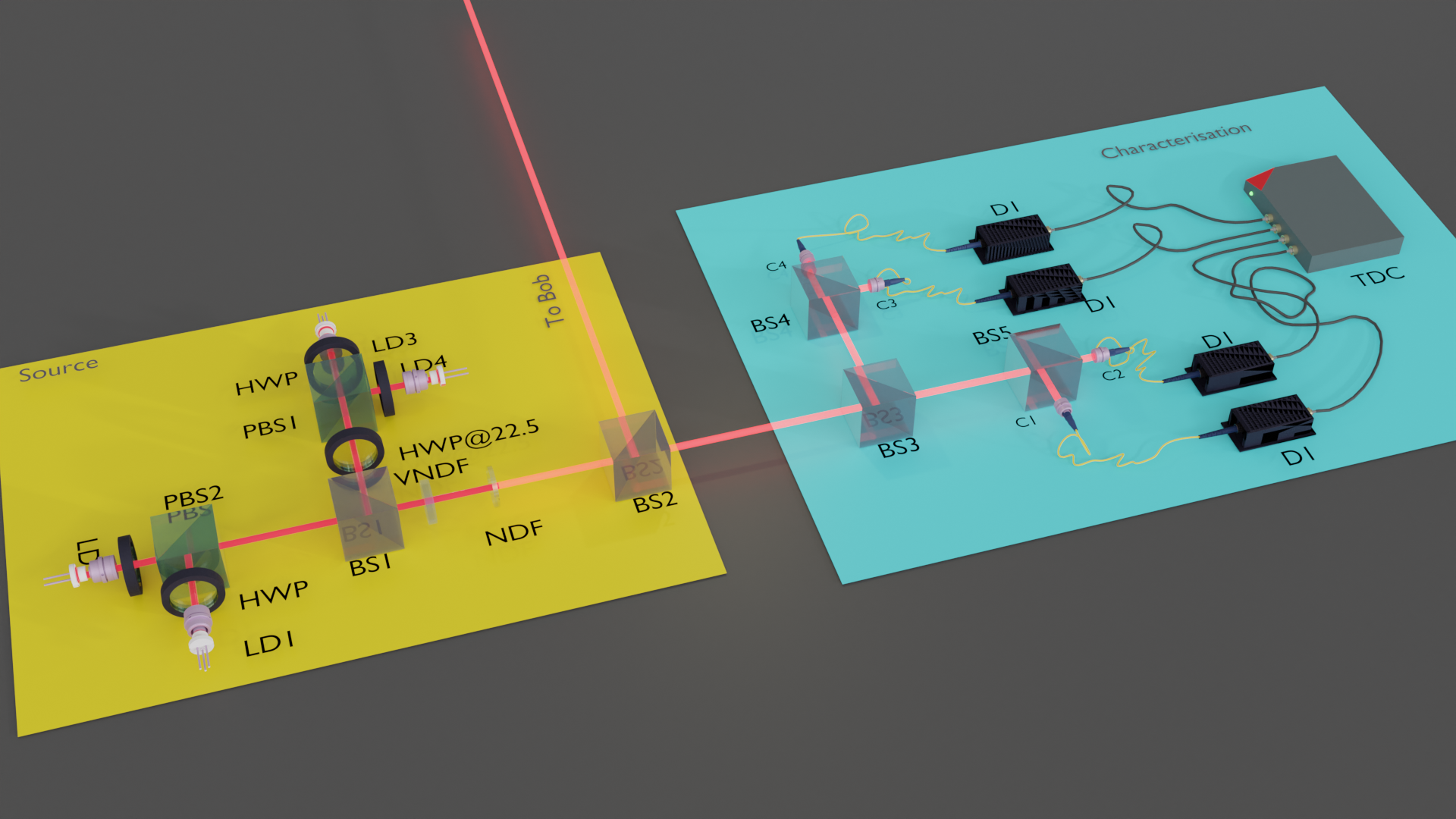}
    \caption{Experimental setup for characterising photon statistics of the source: V-NDF: Variable Neutral Density Filter; NDF: Neutral Density Filter; HWP: Half Wave Plate; BS: Beam Splitter; PBS: Polarizing Beam Splitter; M: Mirror; C: Coupler; D: Single Photon Counting Module; TDC: ID-900 Time Controller.}
    \label{fig:I_BE}
\end{figure*}

The experimental setup comprises four diode lasers of 808 nm (L808P010 Thorlabs). The source setup is the standard BB84 transmitter, using four laser diodes to contribute to the four polarisation states for QKD. The source comprises the laser diodes along with the attenuating optics.

Voltage pulses trigger each laser diode with a repetition rate of 1.25 MHz. A half-wave plate and a polarising beam splitter (PBS) combination direct the emitted beam from laser diodes, functioning as both a polariser and an attenuator. The polarising beam splitter (PBS) transmits the horizontally polarised light and reflects vertically polarised light. The rotation of the half-wave plate determines the intensity of each beam. So far, the setup is identical for both bases. To convert the polarisation to the diagonal basis, we insert a half-wave plate and rotate it by $22.5^\circ$. This adjustment ensures that the beam's polarisation aligns with the diagonal basis. The two beams combine at a beam splitter (BS1), and we only consider the output from one arm, discarding the other. The beam intensity reduces to half due to the combination of the four beams at this beam splitter.

Next, all four beams pass through a variable neutral density filter (NDF) with a maximum optical density (OD) of 4 and a fixed NDF with an optical density of 6. These filters reduce the intensity of the beams accordingly. We will analyse the photon statistics of the resulting beam from this final NDF containing the signal. To characterise the photon statistics of each source individually, we block the other sources and analyse them one by one.

The characterisation setup consists of three beam splitters and four detectors. The detectors utilised in the setup are single photon counting modules(Excelitas-SPCM-AQRH-14). The incoming beam is coupled to the SPADs using fiber couplers (Thorlabs-CFC5-B) and multi-mode fibers (Thorlabs-M42L02). Upon reaching beam splitter BS3, the signal divides into two arms. These arms are further split into two more arms by the beam splitters BS4 and BS5, resulting in four arms coupled to the four detectors.

For each source, we recorded timestamps and recorded two, three, and four-fold coincidences among all four detectors. We rotated the variable NDF to acquire various values of $\mu$. The estimation of $\mu$ was approximately determined using equation \eqref{eq19}.
\begin{equation}\label{eq19}
    N=\mu \nu_{rep}\eta
\end{equation}
Where $N$ is the number of counts in the detector per second, $\mu$ is the mean photon number, $\nu_{rep}$ is the repetition rate, i.e. 1.25 MHz, and $\eta$ is the overall efficiency. As discussed, the overall efficiency given by \eqref{eq10}  consists of the quantum efficiency of detector $65\%$, coupling efficiency and branching efficiency. The beam splitters are not 50-50; we also included those experimental efficiencies in our calculations. We characterized the beam splitters, and Table-\ref{tab1} contains the mentioned transmittance and reflectance values.

\begin{table}
    \centering
    \begin{tabular}{|c|c|c|c|c|c|}
    \hline Beam Splitter& $T^2$ & $R^2$ \\
    \hline \hline
    BS3 & $\ \ 0.494\ \ $ & $\ \ 0.453\ \ $ \\\hline 
    BS4 & $0.474$ & $0.446$ \\\hline 
    BS5 & $0.461$ & $0.456$ \\\hline
    \end{tabular}
    \caption{Transmittance and reflectance of beam-splitters used in characterisation setup}
    \label{tab1}
\end{table}

Thus using \eqref{eq19}, we can roughly estimate if we have achieved the desired value of $\mu$ by looking at the number of counts in a single detector as discussed in \ref{M1} since we know that these counts only signify the presence and absence of a pulse containing photons. We now characterise the source using the r-fold coincidences recorded to estimate $\mu$ using \ref{M2}. We repeat this process for all four laser diodes emitting different polarisations. 

We consider a single detector for each source to study the intensity fluctuations of all four sources. Furthermore, we adjust the variable attenuator by rotating it to attain the desired count rate, and the detectors capture individual signals. This information is documented over multiple cycles to examine the fluctuations in the sources. We repeat this experiment for various source intensities to examine fluctuations as a function of intensity.

\section{\label{sec:RnD}Results and Discussion}

\begin{figure}[htp]
    \centering
    \includegraphics[scale=0.5]{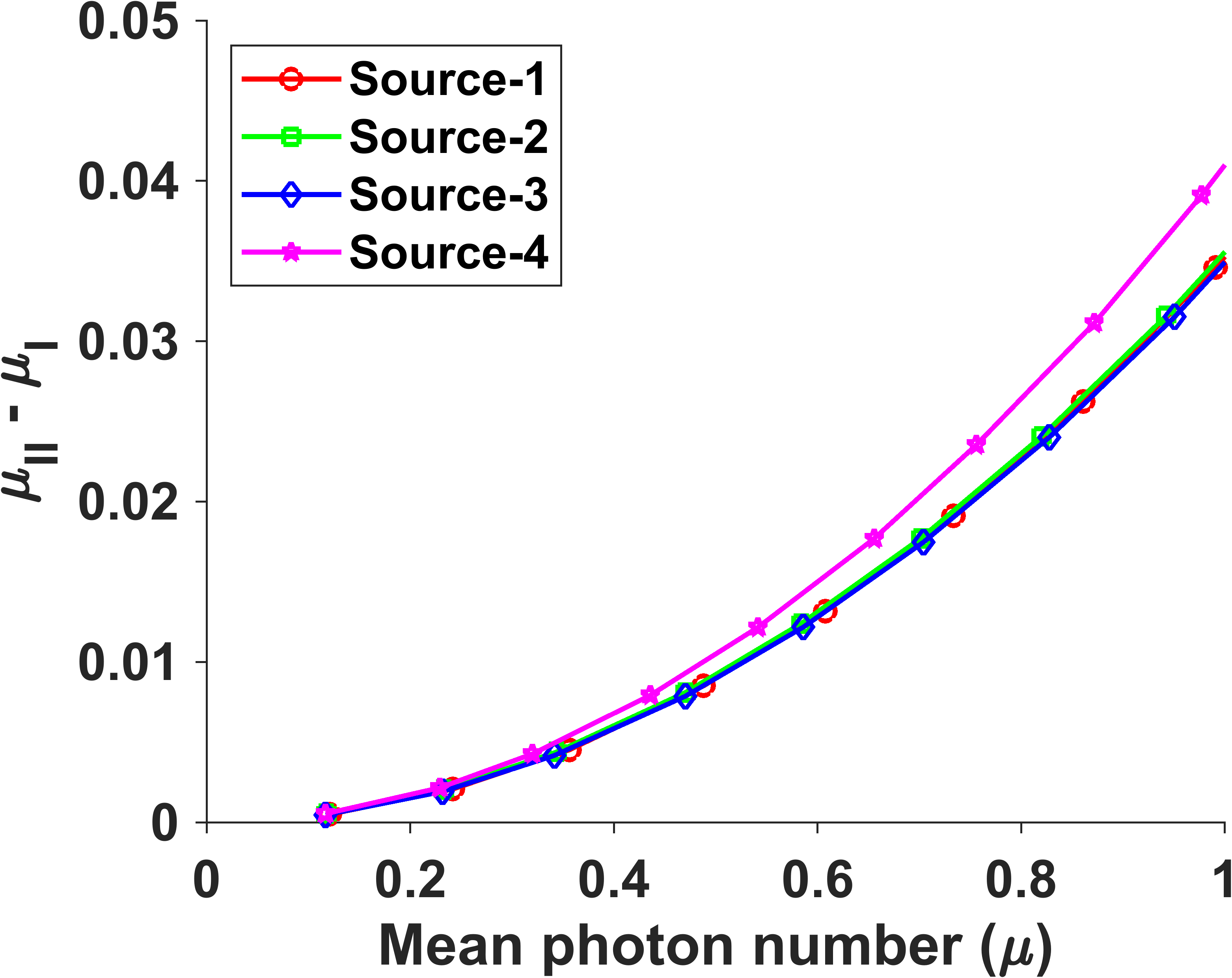}
    \caption{Difference between the mean photon number ($\mu$) calculated using Method-I and Method-II of mean photon number ($\mu$).}
    \label{fig:diffmu}
\end{figure}

We recorded the single and coincidence detections. The photon statistics were verified as Poissonian, as discussed in \ref{M2}. We compared the calculated values from Method-I (\ref{M1}) and Method-II (\ref{M2}), and we plotted the difference as a function of mean photon number ($\mu$) in Figure-\ref{fig:diffmu}. The difference in the estimated values increases with an increase in the mean photon number. This outcome is predictable, as a single detector lacks the accuracy to measure accurately in the presence of multi-photon pulses. Coincidences using multiple on-off detectors enable us to resolve the photon numbers with higher accuracy. 

\begin{figure}[htp]
    \centering
    \includegraphics[scale=0.5]{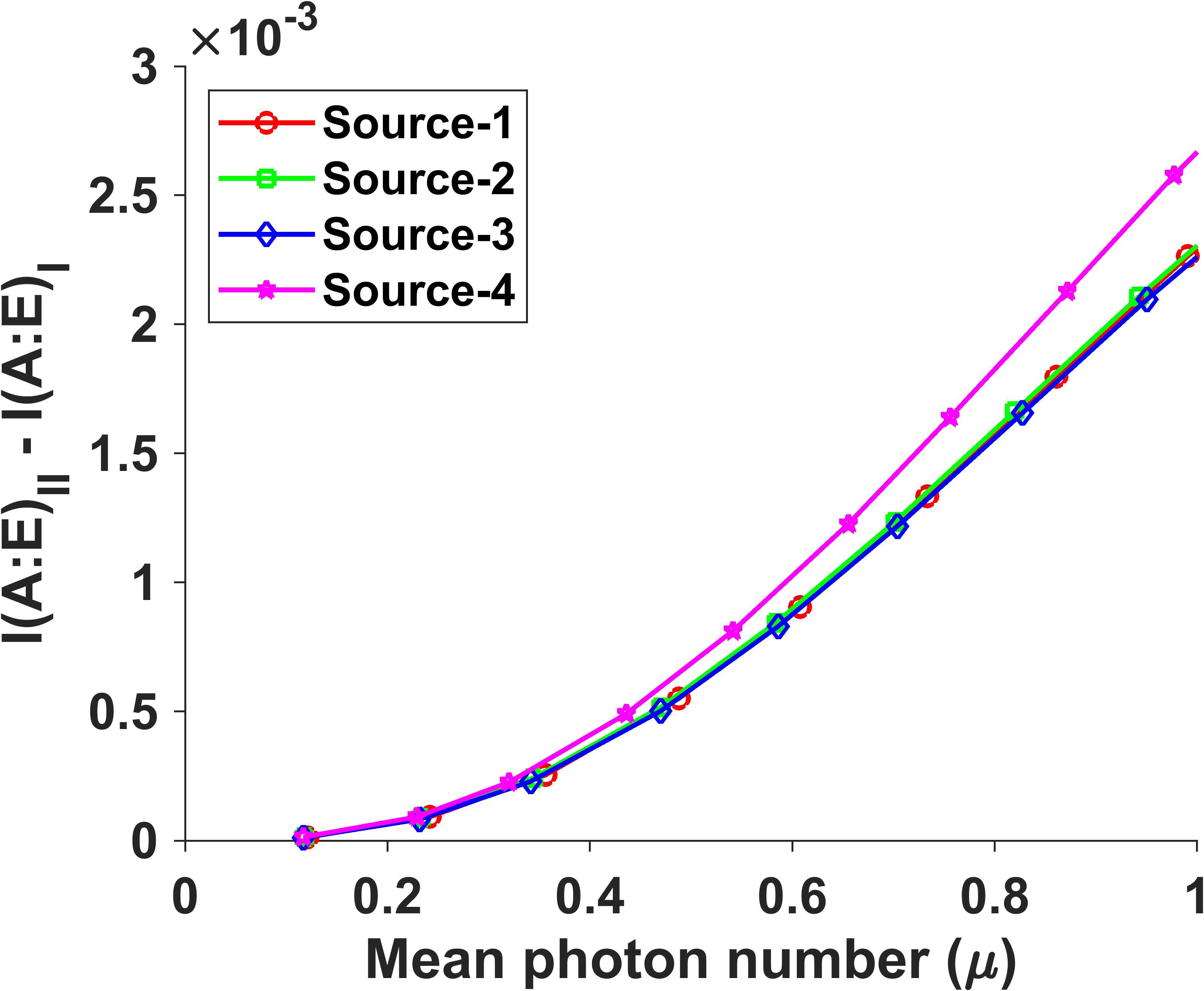}
    \caption{Information leakage due to miscalculated mean photon number ($\mu$) as a function of mean photon number ($\mu$)}
    \label{fig:DIAE}
\end{figure}

We calculate the mutual information for each case using \eqref{eq18}. It's important to note that we are considering the worst-case scenario, where Eve can extract information from all multi-photon pulses in the BB84 protocol. Researchers can extend this study to encompass other prepare-and-measure protocols that utilize WCPs as the source. Accurately estimating potential information leakage is crucial, achievable only through a well-characterized understanding of photon statistics. The errors in mean photon estimation lead to a wrong estimation of I(A:E). An adversary can gain information from the part of the information that goes un-estimated. The corresponding difference in the information leakage I(A:E) as a function of mean photon number($\mu$) can be seen in Figure-\ref{fig:DIAE}

\begin{figure}[htp]
    \centering
    \includegraphics[scale=0.5]{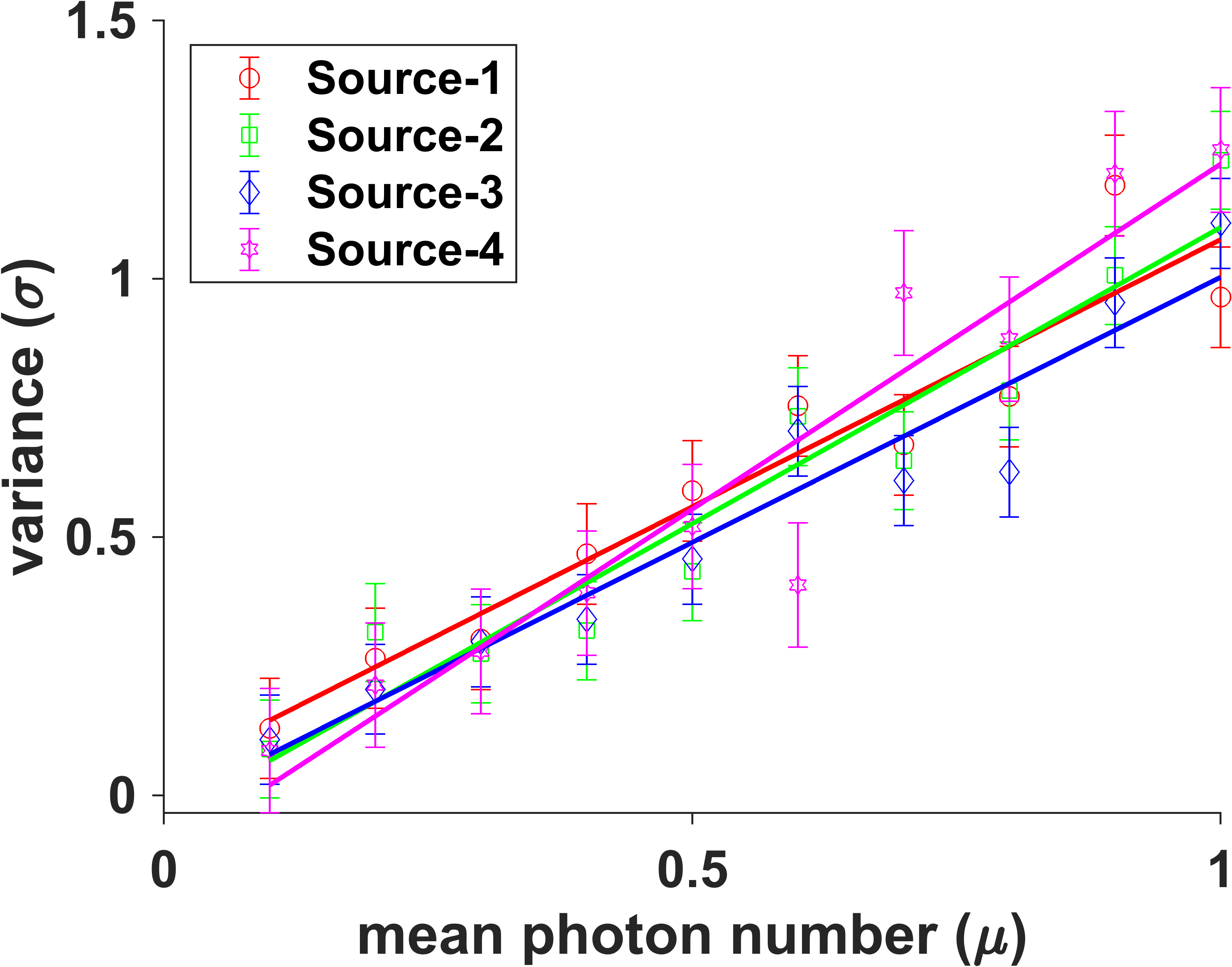}
    \caption{The variation in intensity fluctuations for all four sources vs the average photon count ($\mu$).}
    \label{fig:5} 
\end{figure}

The variation in intensity fluctuations for all four sources relative to the average photon count is illustrated in Figure-\ref{fig:5}. Using a consistent detector, we derive these fluctuations from numerous iterations of single-count data collected from each source. The error bars indicate the disparity between data points and the fitted linear curve. As anticipated, augmenting the mean photon count leads to a rise in intensity fluctuations. 

\begin{figure}[htp]
    \centering
    \includegraphics[scale=0.5]{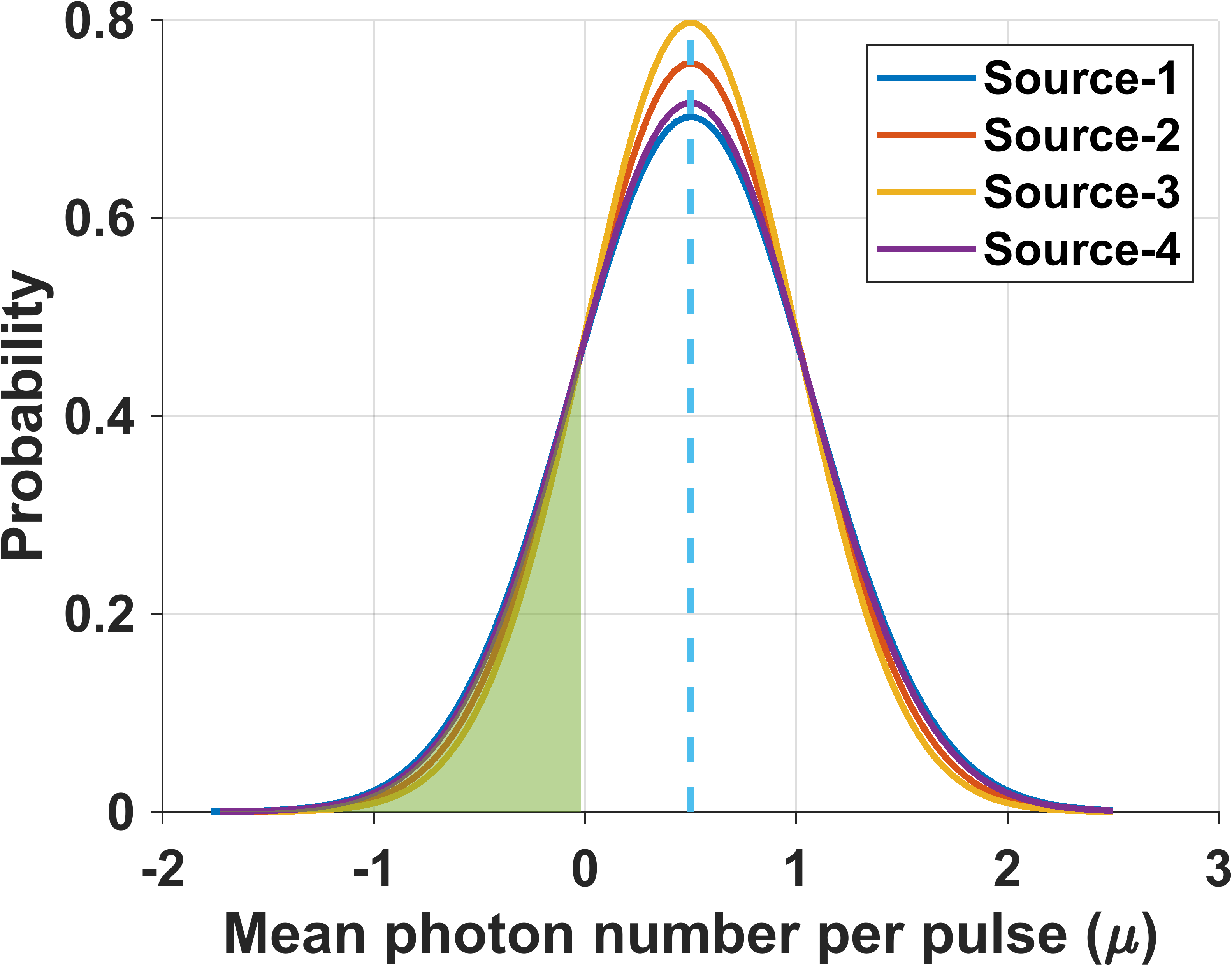}
    \caption{The distribution of all four sources at an average value of $0.5$ photons per pulse}
    \label{fig:6}
\end{figure}

Figure-\ref{fig:6} illustrates the distribution of all four sources at an average value of $0.5$ photons per pulse. Using the fitted data, we create these plots to compare all four sources by analyzing their fluctuations at a specific value of $\mu$. The area of the shaded region signifies the probability of no detection, as the number of counts is never negative.

We report the correlations and potential information leakage in Table-\ref{tab:1}. Ensuring the absence of disparities among the sources and their consistent uniformity is crucial. Varied intensity fluctuations among the sources could empower Eve to extract information, possibly resulting in a side-channel attack. Eve's attempt to gather information between the two bits or across the two bases is the reason we have examined the one-on-one correlations among all the sources. It is thus essential to estimate the amount of information leakage to ensure secure quantum communication.
\begin{table}[htp]
    \centering
    \caption{The correlations $R$ and the potential information leakage $I(A:E)$ among different sources.}
    \label{tab:1}
    \begin{tabular}{|p{2cm}||p{2cm}|p{2cm}|p{2cm}||p{2cm}|}
        \hline
        \hfil \textbf{Sources} & \hfil \textbf{R} & \hfil \textbf{I(A:E)}  \\ \hline \hline
        \hfil S1 $\&$ S2 & \hfil0.9904 & \hfil0.0027 \\ \hline
        \hfil S1 $\&$ S3 & \hfil0.9715 & \hfil0.0082 \\ \hline
        \hfil S1 $\&$ S4 & \hfil0.9993 & \hfil0.0002 \\ \hline
        \hfil S2 $\&$ S3 & \hfil0.9949 & \hfil0.0014 \\ \hline
        \hfil S2 $\&$ S4 & \hfil0.9948 & \hfil0.0014 \\ \hline
        \hfil S3 $\&$ S4 & \hfil0.9796& \hfil0.0058 \\ \hline
    \end{tabular}
\end{table}

\section{ \label{sec:conc}Conclusion}
The weak coherent source used in implementing QKD follows Poisson distribution. The characterisation of this source is essential for accurately estimating any information leakage due to multi-photon pulses. The SPADs used are not photon resolving; hence using multiple on-off detectors offers a better resolution. Four SPADs suffice for the QKD applications as we are well below one photon per pulse on average. The inaccurate measurement of the mean photon number may lead to unaccounted information leakage, leading to an undetected adversarial attack. It is essential to rigorously characterise mean photon numbers in practical QKD systems based on WCPs. The difference in the mean photon number and the corresponding information leakage is estimated. The values increase with an increase in the mean photon number. We can use one detector method for smaller $\mu$ values, i.e. $\mu \leq 0.3$, as, beyond that, the approximation deviates from rigorous characterisation drastically. 

The fluctuations among the four sources exhibit minor variations, resulting in a maximum information leakage on the order of $10^{-2}$ bits per pulse. These source fluctuations escalate as intensity increases, prompting a reconsideration of the intensity discrepancy between the decoy and signal states. Should the difference be substantial, it could expose vulnerabilities to side-channel attacks by Eve. It is essential to analyze to determine the potential for future attacks stemming from these variations in fluctuations.


\section*{Acknowledgement}
The authors acknowledge the partial funding support from DST through the QuST program. The authors are also thankful to group members of the QST lab for their valuable inputs.
\section*{Disclosure}
The authors declare no conflicts of interest.



\bibliography{main}


\end{document}